\documentclass[a4paper,11pt]{article}
\usepackage{jheppub}
\usepackage{bm}
\usepackage{mathtools}

\newcommand{\dd}{\mathrm{d}}

\newcommand{\cV}{\mathcal V}

\title{\boldmath Twisted Cohomology of D-Brane Disk Integrals}

\author{Komeil Babaei Velni}
\affiliation[]{Department of Physics, University of Guilan, P.O. Box 41335-1914, Rasht, Iran}

\emailAdd{babaeivelni@guilan.ac.ir}

\abstract{Disk-level scattering amplitudes involving one Ramond–Ramond and two NSNS states on a D-brane obey nontrivial identities among scalar world-sheet integrals imposed by gauge and T-duality Ward identities. We investigate the cohomological organization of these relations using twisted de Rham cohomology and show that the Ward identities admit twisted-exact representatives. In the $I/J$ sector, the unique generic linear dependence among the Ward identities is realized one degree lower in the same twisted de Rham complex, revealing a common cohomological origin for the identities and their syzygy. We also identify independent pointwise algebraic relations among the integrands. An independent cohomological rank computation then establishes that the physically selected $I/J$ family spans a five-dimensional subspace of the top twisted cohomology at generic complexified kinematics. Extending the construction to the $K$-family, including the exceptional double-pole representative $K_3$, introduces no additional cohomology direction. For a D1-brane amplitude, we exhibit a physical polarization component in which six scalar periods reduce to one at special kinematics.
}


\begin{document}
\maketitle
\flushbottom

\section{Introduction}
\label{sec:intro}

Disk amplitudes provide a fundamental world-sheet description of tree-level
string interactions in the presence of D-branes and have long played an
important role in extracting the corresponding world-volume couplings
\cite{GarousiMyers1996,HashimotoKlebanov1997,BeckerGuoRobbins2012}.
More generally, world-sheet representations of string amplitudes give rise
to families of integrals whose individual expressions can obscure
nontrivial relations among them. In open-string amplitudes, such relations
underlie substantial reductions of disk integrals through partial-fraction,
integration-by-parts, and related identities
\cite{MafraSchlottererStieberger2011,MafraSchlotterer2017}.
For amplitudes involving closed-string insertions on the disk, the boundary
conditions and the interplay between bulk and boundary kinematics introduce
additional structure. This becomes particularly relevant for amplitudes
involving one Ramond--Ramond (RR) and two NSNS states, where distinct scalar
integrals multiply different kinematic structures
\cite{BeckerBeckerRobbinsSu2016,MousaviVelni2018}.

In this class of D-brane amplitudes, gauge invariance and T-duality impose nontrivial relations among the scalar integrals that are needed for the amplitude to organize into the appropriate gauge- and duality-covariant combinations
\cite{GarousiMir2011b,VelniGarousi2013,VelniGarousi2014,BabaeiVelni2016,BabaeiVelniJalali2018, BabaeiVelniJalali2017DBI}.
In previous analyses, these identities were obtained primarily by imposing spacetime gauge Ward identities and T-duality constraints on the computed amplitude. This raises a more structural question: do these spacetime-derived integral relations admit an intrinsic world-sheet
interpretation?

Twisted de Rham cohomology provides a natural framework for addressing this question. In this formulation, the Koba--Nielsen factor defines
the twist, while string amplitudes can be described in terms of pairings of twisted cycles and cocycles \cite{Mizera2017,Mizera2018}. World-sheet total-derivative identities can then be represented by twisted-exact differential forms, providing a direct cohomological interpretation of
relations among the disk integrals. Importantly, this framework also allows relations among such identities to be studied within the same
differential complex.

The aim of this paper is to develop this viewpoint explicitly for the D-brane disk-integral system relevant to one RR and two NSNS states. In addition to constructing world-sheet representatives of the Ward relations, we identify pointwise algebraic relations among the integrands that are independent of the Ward system. Combining these two types of relations makes it possible to determine the independent twisted-cohomology classes spanned by the physically selected integral family. More generally, this organization provides a systematic way to separate integral classes, relations, and redundancies among relations, and may be useful for more complicated D-brane amplitudes with larger world-sheet integral systems.

More concretely, after extracting the common Koba--Nielsen factor \(K_f\),
the disk integrals can be regarded as periods of rational top forms against
a multivalued local system. Defining
\begin{equation}
\omega=d\log K_f,\qquad
\nabla_\omega=d+\omega\wedge ,
\label{eq:twisted-differential-intro}
\end{equation}
world-sheet total derivatives are represented by twisted-exact forms.
Relations among the scalar integrals can then arise in two conceptually distinct ways: as pointwise algebraic identities among their rational representatives, or as cohomological identities generated by \(\nabla_\omega\). This distinction is essential for determining the independent integral content, since relation counting provides an upper bound, whereas minimality requires an independent lower bound on the
dimension of the selected cohomology span.

An independent lower bound on the dimension of the selected cohomology span can be obtained from the critical-locus description of twisted cohomology. For smooth very affine varieties, under suitable genericity conditions, the critical quotient associated with the multivalued function provides an algebraic tool for testing the independence of top-degree twisted-cohomology classes \cite{MatsubaraHeoTelen2025}. We apply this construction to the six-parameter physical kinematic family relevant to the present disk amplitude and establish the required independence through a degeneration-and-scaling argument. The distinction between generic exponent space and the kinematically realized family is important, since additional constraints may arise in special D-brane configurations.

Earlier RNS analyses of the one-RR--two-NSNS disk amplitude organize the world-sheet result in scalar integral families conventionally denoted by $I$, $J$, and $K$ \cite{VelniGarousi2013,MousaviVelni2018}. While relations among these integrals were inferred from NSNS gauge invariance and T-duality, their intrinsic world-sheet realization, the dependencies among the resulting relations, and the minimal dimension of the corresponding twisted-cohomology span were not addressed in those analyses.

Our main result is that the complete \(I/J/K\) family considered here spans
exactly a five-dimensional subspace of the top twisted cohomology at generic
complexified physical kinematics,
\begin{equation}
\dim \operatorname{span}_{H^3}\{I/J/K\}=5 .
\label{eq:five-master-intro}
\end{equation}
A convenient basis for this physically selected subspace is
\begin{equation}
\mathcal B_{\rm C}
=
\{[I_1],[I_2],[I_3],[J_{13}],[J_{14}]\}.
\label{eq:basis-intro}
\end{equation}
Importantly, this statement concerns the cohomological span of the
physically selected integral family and does not imply that the full top
twisted cohomology is five-dimensional.

The proof proceeds in complementary steps. In the $I/J$ sector, the eight Ward identities considered here have generic rank seven and admit explicit twisted-exact representatives. Their unique generic linear dependence is itself generated one degree lower in the same twisted de Rham complex. Two additional relations hold pointwise at the level of
the rational integrands and are independent of the Ward system.
Together, these results give an upper bound of five on the dimension of the selected cohomology span. An independent critical-quotient analysis, supplemented by the corresponding deformation argument, shows that this bound is saturated.

The extension to the \(K\)-family provides a nontrivial test of this structure. The six new \(K\)-dependent Ward identities are again shown to be twisted exact, while four additional relations hold pointwise at the integrand level. Together, these relations generically reduce all \(K\)-representatives except \(K_3\) to the five-dimensional \(I/J\) cohomological sector. The remaining \(K_3\) representative is structurally exceptional because its rational coefficient contains a double divisor pole. We treat it separately by an explicit pole-lowering twisted primitive and show that the resulting simple-pole remainder reduces to the already controlled sector. Thus the enlarged \(I/J/K\) family introduces no additional cohomology direction.

Our use of the explicit scalar-integral representation is
complementary to more compact representations of mixed and
closed-string disk amplitudes in terms of open-string
amplitudes \cite{Stieberger2009, StiebergerTaylor2016, BischofHaack2021, BischofHaackStieberger2023}. The two organizations address different questions: here the $I/J/K$ representation is used to expose relations among the scalar world-sheet integrals and to determine the dimension
of their selected twisted-cohomology span. The resulting
five-dimensional count therefore need not coincide with the number of independent open-string amplitudes appearing in alternative representations.

The remainder of this paper is organized as follows. Section~\ref{sec:family} introduces the disk-integral family and its kinematics. Section~\ref{sec:worldsheet} derives the relevant world-sheet identities, and section~\ref{sec:twisted}  formulates them in twisted de Rham cohomology and analyzes the dependence among the Ward relations.
Section~\ref{sec:counting} discusses relation counting and its consequences. Section~\ref{sec:critical} establishes the independent cohomological rank through the critical-quotient analysis. Section~\ref{sec:Kfamily} extends the construction to the \(K\)-family, including the separate treatment of \(K_3\). Section~\ref{subsec:physical-disk-reduction} applies the reduction to the scalar RR, two-form and graviton amplitude on a D1-brane. We conclude in section~\ref{sec:discussion}.
 
\section{The three-closed-string disk-integral family}
\label{sec:family}
We consider the disk amplitude of three closed-string states in the presence of a D\(p\)-brane, with external momenta \(p_1,p_2,p_3\). For the amplitudes of interest, state \(1\) is the RR state and states \(2\) and \(3\) are NSNS states. The corresponding world-sheet integrals share a common Koba--Nielsen factor and differ in their rational integrands. In this section we fix the kinematic and world-sheet conventions and define the scalar integral representatives used below.

\subsection{Kinematics and common Koba--Nielsen factor}

The presence of the D\(p\)-brane is encoded by the reflection matrix
\(D^\mu{}_\nu\), with \(+1\) and \(-1\) eigenvalues along the world-volume
and transverse directions, respectively. We define the corresponding
projectors
\[
 V=\frac{\eta+D}{2},\qquad N=\frac{\eta-D}{2}.
\]
We use the shorthand
\(p_i V p_j:=p_i^\mu V_{\mu\nu}p_j^\nu\) and
\(p_i N p_j:=p_i^\mu N_{\mu\nu}p_j^\nu\).
The three external momenta are on shell and satisfy momentum conservation
along the D-brane world-volume,
\[
 p_i^2=0,\qquad V\cdot(p_1+p_2+p_3)=0.
\]
Let \(z_i\) denote the insertion points of the three closed-string vertices
on the disk. After fixing the disk conformal Killing group, we choose
\[
z_1=0,\qquad
z_2=x,\qquad
z_3=y e^{i\theta},
\qquad
0<x,y<1,\quad 0<\theta<2\pi .
\label{eq:disk-gauge-fixing}
\]
With this parametrization, the bosonic world-sheet correlator gives a common Koba--Nielsen factor for the scalar integrals considered below,
\begin{equation}
K_f
=
x^{\alpha}y^{a}(1-x^2)^{e}(1-y^2)^{d}
\Delta^{b}\Sigma^{c},
\label{eq:Kf}
\end{equation}
where
\[
\Delta=x^2+y^2-2xy\cos\theta,
\qquad
\Sigma=1+x^2y^2-2xy\cos\theta .
\]
The six exponents are determined by the closed-string kinematics,
\begin{equation}
\begin{aligned}
\alpha &=2p_1\!\cdot p_2,
&\qquad
a &=2p_1\!\cdot p_3,
&\qquad
b &=p_2\!\cdot p_3,
\\
c &=p_2\!\cdot D\!\cdot p_3,
&
d &=p_3\!\cdot D\!\cdot p_3,
&
e &=p_2\!\cdot D\!\cdot p_2 .
\end{aligned}
\label{eq:KN-exponents}
\end{equation}
These parameters are inherited from the physical momentum invariants and
should therefore not be regarded as a priori independent exponents.

This Koba--Nielsen factor is the one obtained in the RNS analysis of the corresponding three-closed-string disk amplitudes; see, for example,
refs.~\cite{GarousiMir2011b,VelniGarousi2013,MousaviVelni2018}.

The scalar integrals in this representation can therefore be written in the uniform form
\begin{equation}
I_i=\int_{\Gamma_{\rm disk}} K_f\,\phi_i ,
\label{eq:general-period}
\end{equation}
where \(\Gamma_{\rm disk}=(0,1)_x\times(0,1)_y\times S^1_\theta\)
is the physical integration cycle and \(\phi_i\) is the corresponding rational three-form. For example, the convention adopted here for the basic integral \(I_1\) is
\begin{equation}
 I_1=\int_0^1\dd x\int_0^1\dd y\int_0^{2\pi}\dd\theta\,
 \frac{K_f}{xy}.                                                       \label{eq:I1}
\end{equation}

\subsection{A logarithmic representation}

For the cohomological analysis it is convenient to rewrite the angular
dependence in logarithmic form. We therefore introduce
\[
z=e^{i\theta},
\qquad
d\theta=\frac{1}{i}\,d\log z .
\]
In terms of \(z\), the two angular factors become
\[
\Delta=\frac{(x-yz)(xz-y)}{z},
\qquad
\Sigma=\frac{(1-xyz)(z-xy)}{z}.
\label{eq:DeltaSigma-factorized}
\]

Their ratio occurs naturally in the logarithmic representatives below, so
we define
\begin{equation}
Q:=\frac{\Delta}{\Sigma}
=
\frac{(x-yz)(xz-y)}
     {(1-xyz)(z-xy)} .
\label{eq:Q}
\end{equation}

The \(I\)-sector rational three-forms introduced above admit particularly simple logarithmic representatives:
\begin{equation}
\begin{aligned}
\phi_1 &=
 \frac{1}{i}\,
 d\log x\wedge d\log y\wedge d\log z,
\\
\phi_2 &=
 \frac{1}{i}\,
 d\log x\wedge d\log Q\wedge d\log z,
\\
\phi_3 &=
 \frac{1}{i}\,
 d\log Q\wedge d\log y\wedge d\log z,
\\
\phi_4 &=
-\frac{1}{i}\,
 d\log x\wedge
 d\log\!\left(\frac{y}{1-y^2}\right)
 \wedge d\log z,
\\
\phi_7 &=
-\frac{1}{i}\,
 d\log\!\left(\frac{x}{1-x^2}\right)
 \wedge d\log y\wedge d\log z .
\end{aligned}
\label{eq:I-log-representatives}
\end{equation}
For any scalar integral $A$ in the disk-integral family, we denote its rational prefactor by $R_A$, so that the corresponding integrand is written as
\begin{equation}
K_f R_A\,dx\wedge dy\wedge d\theta .
\label{eq:rational-prefactor}
\end{equation}
For the representatives above, for example,
\[
R_{I_2}=\frac{1}{x}\,\partial_y\log Q,
\qquad
R_{I_3}=\frac{1}{y}\,\partial_x\log Q.
\]
These logarithmic expressions are algebraically equivalent to the original disk integrands. For reference, the explicit rational representatives of the \(I/J/K\) families used throughout the analysis are collected in appendix \ref{app:rational-representatives}.

\subsection{The Ward-identity system}

The RNS amplitude analysis leads to eight Ward relations among the fourteen
scalar integrals
\[
\{I_1,I_2,I_3,I_4,I_7,
J,J_1,J_2,J_3,J_4,J_5,J_{12},J_{13},J_{14}\}.
\]
These relations are required for the amplitude to organize into
gauge-invariant and T-duality-compatible combinations \cite{VelniGarousi2013,VelniGarousi2014}. For example, one of the \(I\)-sector Ward relations is
\begin{equation}
 2(p_1Np_3)I_1+2(p_3Vp_3)I_4
 +(p_2Np_3)I_2-(p_2Vp_3)I_3=0,
 \label{eq:Iward}
\end{equation}
in the convention for \(I_1\) adopted in eq.~\eqref{eq:I1}.
This convention differs by an overall sign from that used for the
corresponding basic integral in the earlier amplitude analysis.

The question is therefore whether these relations can be derived directly at the level of the world-sheet integrals, rather than inferred only after imposing the spacetime Ward identities.
\section{Integral identities from the world sheet}
\label{sec:worldsheet}
We now address this question directly at the level of the world-sheet integrands. Two distinct mechanisms arise. Some relations hold pointwise as algebraic identities among the rational integrands, whereas others are generated by total derivatives on the disk integration space. The latter will provide direct world-sheet representatives of the Ward identities introduced in the previous section.
\subsection{Pointwise identities}
Before turning to total-derivative relations, we first isolate two identities that hold pointwise at the level of the rational integrands.
In the conventions adopted above, they are
\begin{align}
J_{12}-J_2-2J_5 &=0,
\label{eq:pointwise-1}\\
-I_1+J+J_2+J_5 &=0.
\label{eq:pointwise-2}
\end{align}
These relations are stronger than identities obtained only after integration. In the notation of eq.~\eqref{eq:rational-prefactor}, these relations amount to the pointwise identities
\[
R_{J_{12}}-R_{J_2}-2R_{J_5}\equiv0,
\qquad
-R_{I_1}+R_J+R_{J_2}+R_{J_5}\equiv0.
\]
Thus the cancellation occurs before multiplication by the common
Koba--Nielsen factor is integrated over \(\Gamma_{\rm disk}\).
Equivalently, after bringing the rational functions to a common
denominator, the corresponding numerators vanish identically.

Consequently, these identities do not rely on world-sheet total
derivatives, the choice of integration cycle, or the treatment of boundary terms.
This distinguishes them from the Ward relations considered below,
whose world-sheet realization involves total derivatives.

\subsection{A radial total-derivative identity}

To reproduce the representative I-sector Ward relation in
eq.~\eqref{eq:Iward}, consider the radial total derivative
\begin{equation}
 \int_{\Gamma_{\rm disk}}
 \dd x\,\dd y\,\dd\theta\,
 \partial_y\!\left(\frac{K_f}{x}\right)=0 .
 \label{eq:radialIBP}
\end{equation}
The equality is understood first in a domain of complexified kinematics in which the integral is absolutely convergent and the endpoint contribution vanishes, followed by meromorphic continuation. This is the standard analytic prescription for total-derivative identities in Euler-type string integrals.

The logarithmic derivative of the Koba--Nielsen factor is
\begin{equation}
y\partial_y\log K_f
=
a-\frac{2d\,y^2}{1-y^2}
+b\,y\partial_y\log\Delta
+c\,y\partial_y\log\Sigma .
\label{eq:y-log-Kf}
\end{equation}
We also use the elementary identities
\[
x\partial_x\log\Delta+y\partial_y\log\Delta=2,
\qquad
x\partial_x\log\Sigma=y\partial_y\log\Sigma .
\]
Substitution into eq.~\eqref{eq:radialIBP} gives
\begin{equation}
(a+b+c+d)I_1
+\frac{b-c}{2}I_2
-\frac{b+c}{2}I_3
+dI_4=0 .
\label{eq:radial-integral-relation}
\end{equation}
Using the on-shell conditions and momentum conservation parallel to the brane, the coefficients can be rewritten as
\[
\begin{aligned}
a+b+c+d &=2(p_1Np_3),
&\qquad d&=2(p_3Vp_3),\\
\frac{b-c}{2}&=p_2Np_3,
&
-\frac{b+c}{2}&=-p_2Vp_3 .
\end{aligned}
\]
Substituting these identities into eq.~\eqref{eq:radial-integral-relation} reproduces exactly the representative Ward relation in eq.~\eqref{eq:Iward}. The relation obtained by exchanging
\(2\leftrightarrow3\) follows analogously from the corresponding
\(x\)-derivative.

This establishes, for the representative \(I\)-sector relation, that the Ward identity is not merely a spacetime constraint on the integrated amplitude but is generated directly by a world-sheet total derivative.

\subsection{Mixed radial--angular total derivatives}
\label{subsec:Mixed radial}

The remaining nontrivial Ward relations can be generated by total derivatives involving both radial and angular components. We introduce two rational vector fields with components
\[
\begin{aligned}
F_x^{(1)}&=-xR_{I_2},
&\qquad
F_\theta^{(1)}
&=2\sin\theta\,\frac{\Delta+\Sigma}{\Delta\Sigma},
\\
F_x^{(2)}&=xR_{I_3},
&
F_\theta^{(2)}
&=2\sin\theta\,\frac{(x^2-1)(y^2-1)}{\Delta\Sigma}.
\end{aligned}
\]
With these choices, the corresponding weighted divergences
\[
\partial_x\!\left(K_f F_x^{(r)}\right)
+\partial_\theta\!\left(K_f F_\theta^{(r)}\right),
\qquad r=1,2,
\]
can be expressed entirely in the rational integral basis.
After division by the common factor \(K_f\), the resulting rational identities yield, upon integration over \(\Gamma_{\rm disk}\),
\begin{align}
0={}&-2(p_1Np_2)I_2
 +(p_2Np_3)(J_{13}-J_{14})
 +2(p_2Vp_2)J_2 \nonumber\\
&+(p_2Vp_3)(-4J+J_{13}+J_{14}-2J_5),                                 \label{eq:mixed1}\\
0={}&2(p_1Np_2)I_3
 -2(p_2Vp_2)J_1
 +(p_2Vp_3)(J_{13}-J_{14})\nonumber\\
&+(p_2Np_3)(J_{13}+J_{14}+2J_5).                                     \label{eq:mixed2}
\end{align}

The remaining independent relation in this sector is generated by the purely radial total derivative
\begin{equation}
\int_{\Gamma_{\rm disk}} dx\,dy\,d\theta\,
\partial_x\!\left[
K_f\,\frac{1+y^2}{y(1-y^2)}
\right]=0 .
\label{prtd}
\end{equation}
Expanding the derivative and expressing the resulting rational
integrands in the same scalar-integral basis gives
\begin{equation}
-2(p_1Np_2)I_4
+(p_2Np_3)J_{12}
+2(p_2Vp_2)J_3
-(p_2Vp_3)J_4=0 .
\label{eq:mixed3}
\end{equation}
The earlier amplitude analysis also displayed a fifth relation in this sector, quadratic in the kinematic invariants,
\begin{align}
0={}&
(-J_{13}+J_{14})
\Big[(p_2Np_3)^2+(p_2Vp_3)^2\Big]
\nonumber\\
&+2(p_1Np_2)
\Big[I_2(p_2Np_3)-I_3(p_2Vp_3)\Big]
\nonumber\\
&+2(p_2Vp_2)
\Big[(p_2Vp_3)J_1-(p_2Np_3)J_2\Big]
\nonumber\\
&-2(-2J+J_{13}+J_{14})
(p_2Vp_3)(p_2Np_3).
\end{align}
This relation is not an independent Ward identity. It is obtained identically by multiplying eq.\eqref{eq:mixed1} by $-(p_2Np_3)$, eq.\eqref{eq:mixed2} by $-(p_2Vp_3)$, and adding the two resulting equations. Its $2\leftrightarrow3$ image is redundant for the same reason. Consequently, the four linear relations generated here, eqs.~\eqref{eq:radial-integral-relation}, \eqref{eq:mixed1}, \eqref{eq:mixed2}, and \eqref{eq:mixed3}, together with their $2\leftrightarrow3$ images, constitute the eight Ward identities considered in the following analysis.

 These eight linear Ward identities are realized directly at the integrand level: after subtracting the corresponding weighted divergence
from each Ward-integrand combination and bringing the result to a common
denominator, the residual numerator vanishes identically. These results provide explicit world-sheet representatives for the complete Ward system.

\section{Twisted de Rham structure}
\label{sec:twisted}
The total-derivative identities constructed in the previous section admit a natural formulation in twisted de Rham cohomology. We now recast these world-sheet relations in that language and show that the Ward identities are represented by twisted-exact forms.
\subsection{The twisted differential}
\label{subsec:The twisted differential}

Recall from eq.~\eqref{eq:twisted-differential-intro} that the common
Koba--Nielsen factor defines
\[
\omega=d\log K_f,
\qquad
\nabla_\omega=d+\omega\wedge .
\]

In the complex coordinate \(z=e^{i\theta}\), the logarithmic one-form takes the explicit form
\begin{align}
\omega={}&
 \alpha\,\dd\log x+a\,\dd\log y
 +e\,\dd\log(1-x^2)+d\,\dd\log(1-y^2)\nonumber\\
&+b\,\dd\log(x-yz)+b\,\dd\log(xz-y)
 +c\,\dd\log(1-xyz)+c\,\dd\log(z-xy)\nonumber\\
&-(b+c)\,\dd\log z .                                                  \label{eq:omega}
\end{align}
For any differential form \(\eta\),
\[
K_f\nabla_\omega\eta=d(K_f\eta).
\]
Thus a total derivative of the type encountered in section~\ref{sec:worldsheet} is represented, after stripping off the common Koba--Nielsen factor, by a \(\nabla_\omega\)-exact form.

As a first example, consider the radial identity in eq.~\eqref{eq:radial-integral-relation}. Its total-derivative representative is encoded by the
two-form
\[
\eta_y
=
-\frac{1}{i}\,d\log x\wedge d\log z .
\]
This primitive provides the explicit representative that will be denoted by $\tilde{\eta}_0$ in the complete Ward system below.

Acting with the twisted differential gives
\[
\nabla_\omega\eta_y
=
2(p_1Np_3)\phi_1
+(p_2Np_3)\phi_2
-(p_2Vp_3)\phi_3
+2(p_3Vp_3)\phi_4 .
\]
The right-hand side is precisely the three-form combination associated
with the corresponding Ward relation. Hence this Ward combination is
\(\nabla_\omega\)-exact.

The mixed total derivatives of section~3 admit the same
differential-form description. For a vector field with components
\((F_x,F_y,F_\theta)\), define
\begin{equation}
\eta
=
F_x\,dy\wedge d\theta
-F_y\,dx\wedge d\theta
+F_\theta\,dx\wedge dy .
\label{eq:general-two-form}
\end{equation}
Then $d(K_f\eta)$ is the three-form corresponding to the weighted divergence of the associated world-sheet vector field.

\subsection{Primitives for the Ward system}
\label{subsec:Primitives}
We now construct explicit two-form primitives for the eight Ward identities identified in section~\ref{sec:worldsheet}.
We first consider the unexchanged sector. The two-form primitives associated with the purely radial total derivatives may be chosen as
\[
\eta_0=\frac{1}{y}\,dy\wedge d\theta,
\qquad
\eta_3=\frac{1+y^2}{y(1-y^2)}\,dy\wedge d\theta.
\]
In particular, $\eta_3$ corresponds through eq.~\eqref{eq:general-two-form} to the radial total derivative in eq.~\eqref{prtd}.

The remaining two primitives in this sector are the differential-form representatives of the mixed radial--angular total derivatives constructed
in section~\ref{subsec:Mixed radial}:
\[
\eta_1
=
-xR_{I_2}\,dy\wedge d\theta
+
2\sin\theta\,\frac{\Delta+\Sigma}{\Delta\Sigma}\,dx\wedge dy,
\]
\[
\eta_2
=
xR_{I_3}\,dy\wedge d\theta
+
2\sin\theta\,
\frac{(x^2-1)(y^2-1)}{\Delta\Sigma}\,dx\wedge dy.
\]
Through eq.~\eqref{eq:general-two-form}, $\eta_1$ and $\eta_2$ are the two-form primitives associated with the Ward relations in eqs.~\eqref{eq:mixed1} and \eqref{eq:mixed2}, respectively.

The four primitives in the exchanged sector are obtained by applying the $2\leftrightarrow3$ exchange to the corresponding world-sheet total-derivative generators. They may be written as
\[
\widetilde{\eta}_0
=
-\frac{1}{x}\,dx\wedge d\theta,
\qquad
\widetilde{\eta}_3
=
-\frac{1+x^2}{x(1-x^2)}\,dx\wedge d\theta,
\]
and
\[
\widetilde{\eta}_1
=
yR_{I_3}\,dx\wedge d\theta
+
2\sin\theta\,\frac{\Delta+\Sigma}{\Delta\Sigma}\,dx\wedge dy,
\]
\[
\widetilde{\eta}_2
=
-yR_{I_2}\,dx\wedge d\theta
+
2\sin\theta\,
\frac{(x^2-1)(y^2-1)}{\Delta\Sigma}\,dx\wedge dy.
\]
These are the two-form primitives for the $2\leftrightarrow3$ images of the four Ward relations represented by
$\eta_0,\eta_1,\eta_2,\eta_3$, respectively.

We denote the corresponding twisted-exact three-forms by
\[
\Phi_a:=\nabla_\omega\eta_a,
\qquad
\widetilde{\Phi}_a:=\nabla_\omega\widetilde{\eta}_a,
\qquad a=0,1,2,3.
\]
Here $\Phi_1$, $\Phi_2$, and $\Phi_3$ represent the Ward relations in eqs.~\eqref{eq:mixed1} and \eqref{eq:mixed2}, and \eqref{eq:mixed3}, respectively, while $\Phi_0$ represents the $2\leftrightarrow3$ image of eq.~\eqref{eq:radial-integral-relation}. Their exchanged counterparts are represented by
$\widetilde{\Phi}_a$; in particular, $\widetilde{\Phi}_0$
represents eq.~\eqref{eq:radial-integral-relation}, in agreement with the explicit construction in section~\ref{subsec:The twisted differential}.

Hence all eight Ward combinations have explicit representatives in
\[
\operatorname{im}
\bigl(
\nabla_\omega:\Omega^2(X)\longrightarrow\Omega^3(X)
\bigr).
\]
Moreover, their explicit two-form preimages allow the linear dependence among the eight Ward representatives to be lifted one degree lower in the twisted complex, as we now show.


\subsection{The Ward syzygy}
\label{syzygy}

At generic complexified kinematics, the eight Ward relations considered above have rank seven and therefore possess a unique linear dependence, up to an overall normalization. For compactness, we introduce the physical kinematic invariants
\begin{equation}
\begin{aligned}
s &= p_1 N p_2, \qquad t = p_1 N p_3, \qquad u = p_2 V p_3,\\
p &= p_2 V p_2, \qquad q = p_3 V p_3, \qquad v = p_2 N p_3.
\end{aligned}
\label{mandelstum}
\end{equation}

In terms of the twisted-exact three-forms introduced in section~\ref{subsec:Primitives}, this dependence takes the form  
\begin{equation}
0=
2t\,\Phi_0
-v\,\Phi_1
-u\,\Phi_2
-2q\,\Phi_3
-2s\,\widetilde{\Phi}_0
+v\,\widetilde{\Phi}_1
+u\,\widetilde{\Phi}_2
+2p\,\widetilde{\Phi}_3 .
\label{eq:twisted-exact1}
\end{equation}
Equation~\eqref{eq:twisted-exact1} makes the redundancy of the eight Ward representatives explicit at the level of twisted-exact three-forms. More importantly, this dependence is itself generated within the same twisted de Rham complex. The corresponding linear combination of two-form primitives is not merely twisted closed, but twisted exact:
\begin{equation}
\begin{aligned}
&
2t\,\eta_0
-v\,\eta_1
-u\,\eta_2
-2q\,\eta_3
-2s\,\widetilde{\eta}_0
+v\,\widetilde{\eta}_1
+u\,\widetilde{\eta}_2
+2p\,\widetilde{\eta}_3
\\
&\hspace{4cm}
=\nabla_\omega(d\theta).
\end{aligned}
\label{eq:twisted closed1}
\end{equation}

Since
\[
d\theta=\frac{1}{i}\,d\log z,
\qquad
\nabla_\omega(d\theta)=\omega\wedge d\theta ,
\label{eq:dtheta-log}
\]
the one-form generating the lifted syzygy is itself logarithmic on the complexified configuration space. Thus the redundancy among the Ward three-forms is generated one degree
lower within the same twisted de Rham complex. 

The lifted identity can be verified directly using the kinematic relations
\[
\begin{aligned}
2p_1\cdot p_2 &= 2(s-p-u),&
2p_1\cdot p_3 &= 2(t-q-u),\\
p_2Dp_2 &= 2p,&
p_3Dp_3 &= 2q,\\
p_2\cdot p_3 &= u+v,&
p_2Dp_3 &= u-v .
\end{aligned}
\]
Substituting the explicit primitives into the left-hand side of
eq.~\eqref{eq:twisted closed1}, the $dx\wedge dy$ component cancels identically, while the
remaining terms reduce to
\[
\left[
\frac{2s}{x}
+v yR_{I_3}
-u yR_{I_2}
-2p\frac{1+x^2}{x(1-x^2)}
\right]dx\wedge d\theta
\]
\[
+
\left[
\frac{2t}{y}
+v xR_{I_2}
-u xR_{I_3}
-2q\frac{1+y^2}{y(1-y^2)}
\right]dy\wedge d\theta .
\]
Using the explicit logarithmic derivatives of $K_f$, the two coefficients  agree with $\partial_x\log K_f$ and $\partial_y\log K_f$, respectively. Hence the expression above is $\omega\wedge d\theta=\nabla_\omega(d\theta)$, proving eq.~\eqref{eq:twisted closed1} directly at the level of world-sheet differential forms.

Applying $\nabla_\omega$ to eq.~\eqref{eq:twisted closed1} and using $\nabla_\omega^2=0$ gives eq.~\eqref{eq:twisted-exact1}. The logarithmic one-form $d\theta$ thus generates the Ward syzygy through the sequence
\[
\Omega^1(X)
\xrightarrow{\ \nabla_\omega\ }
\Omega^2(X)
\xrightarrow{\ \nabla_\omega\ }
\Omega^3(X),
\]
where $X$ denotes the complement of the singular divisor of the complexified world-sheet integrand.

\section{Relation counting and predictive consequence}
\label{sec:counting}

The cohomological analysis of the previous section determines the independent content of the Ward system: the eight linear Ward generators have a unique generic syzygy and therefore span only seven independent constraints. We now combine this result with the pointwise world-sheet identities derived in section~\ref{sec:worldsheet} to obtain an upper bound
on the dimension of the integral space selected by the amplitude.
This counting also separates the relations implied by the Ward system from additional world-sheet identities carrying genuinely new information.

Let $\cV_{14}$  denote the subspace of
$H^3(X_{K_{\rm phys}},\nabla_\omega)$
spanned by the fourteen selected classes,
\begin{equation}
\cV_{14}
=
\operatorname{span}_{K_{\rm phys}}
\{[I_1],[I_2],[I_3],[I_4],[I_7],
[J],[J_1],[J_2],[J_3],[J_4],[J_5],
[J_{12}],[J_{13}],[J_{14}]\}.
\label{eq:v14}
\end{equation}


Let $R_{\rm Ward}$ denote the coefficient matrix of the eight linear Ward generators with respect to the ordered integral family in eq.~\eqref{eq:v14}. As established in section~\ref{syzygy}, its generic rank is
\begin{equation}
\operatorname{rank}R_{\rm Ward}=7.
\label{eq:ward-rank}
\end{equation}

Appending the two pointwise identities \eqref{eq:pointwise-1} and \eqref{eq:pointwise-2} gives a $10\times14$ relation matrix $R_{\rm ver}$. The Ward syzygy allows the row corresponding to $\Phi_0$ to be omitted without changing the generic rank. The resulting $9\times14$ matrix has rows ordered as \[
\Phi_1,\Phi_2,\Phi_3, \widetilde{\Phi}_0,\widetilde{\Phi}_1, \widetilde{\Phi}_2,\widetilde{\Phi}_3, \] followed by the two pointwise identities.

Selecting the columns \[ I_1,I_2,I_3,I_4,I_7,J,J_1,J_2,J_3 \] in this order gives a $9\times9$ minor $M_9$ whose determinant is
\begin{equation}
\det M_9=256p^2qt^3u.
\label{eq:rank-nine-minor}
\end{equation}
Since this determinant is nonzero at generic kinematics, while the Ward syzygy bounds the rank from above by nine, we obtain
\begin{equation}
\operatorname{rank}R_{\rm ver}=9.
\label{eq:verified-rank-nine}
\end{equation}
Thus the two pointwise identities contribute two independent relation directions beyond the rank-seven Ward system.

Consequently, the fourteen selected classes satisfy nine independent relations, giving
\begin{equation}
\dim_{K_{\rm phys}}\cV_{14}
\leq 14-\operatorname{rank}R_{\rm ver}
=5.
\label{eq:five-upper-bound}
\end{equation}
This is an upper bound: relation counting alone does not establish that five classes remain independent. The complementary lower bound is proved in section~\ref{sec:critical}.

It is useful to compare this counting with the earlier amplitude analysis of \cite{VelniGarousi2014}. There, five relations were presented in each exchange sector, with one quadratic relation that is algebraically generated by two linear ones. We retain four linear Ward generators and their images under $2\leftrightarrow3$. At generic kinematics, the syzygy exhibited in section~\ref{syzygy} is the only independent relation among these eight generators, so the Ward matrix has rank seven. The two pointwise world-sheet identities derived in section~\ref{sec:worldsheet} are linearly independent modulo the span of the Ward generators and therefore raise the rank of the combined relation matrix to nine.


Unlike the Ward relations, the two additional constraints arise from pointwise identities among the world-sheet integrands rather than from further spacetime gauge or T-duality conditions. They therefore expose redundancies of the scalar-integral representation that are not apparent from Ward-constraint counting alone.

Bischof, Haack and Stieberger~\cite{BischofHaackStieberger2023} express the three-closed-string disk amplitude in terms of six-point open-string amplitudes. The bound derived here concerns the dimension of the scalar-integral space $\cV_{14}$ in the $I/J$ decomposition.

\section{Critical quotient and exact master count}
\label{sec:critical}

The rank-nine relation module of section~\ref{sec:counting} provides an upper bound on the dimension of the $I/J$ span, but does not by itself establish that five independent classes remain. To prove minimality independently, we now use
the critical-point degeneration of twisted cohomology. The purpose of this section is therefore to obtain a complementary lower bound and thereby turn the relation-counting result into an exact master count.

Let
\begin{equation}
K_{\rm phys}=\mathbb{C}(\alpha,a,e,d,b,c)
\end{equation}
be the function field generated by the six complexified kinematic exponents entering the disk kernel. The logarithmic one-form $\omega=d\log K_f$ depends linearly on these parameters, and we regard the twisted complex from this point onward as defined over $K_{\rm phys}$. The divisor complement $X$ of the complexified world-sheet integrand, introduced in section~\ref{syzygy}, is a smooth very affine threefold. Its Euler
characteristic can be computed directly by projecting onto the $(x,y)$ variables. The base is
\[
B=(\mathbb C^\ast\setminus\{\pm1\})^2,
\qquad
\chi(B)=4.
\]
For generic $(x,y)\in B$, the $z$-fiber is $\mathbb{C}^\ast$ with the
four points
\[
\frac{x}{y},\qquad \frac{y}{x},\qquad \frac{1}{xy},\qquad xy
\]
removed, and hence has Euler characteristic $-4$. The loci $x=\pm y$ and $xy=\pm1$ correspond, respectively, to the coincidences $x/y=y/x$ and $xy=1/(xy)$. Each coincidence increases the Euler characteristic of the fiber by one. At the four intersections of these loci, both coincidences occur at distinct values of $z$, so the two corrections add. Each of the two loci has Euler characteristic $-4$, and additivity gives
\[
\chi(X)
=
-4\,\chi(B)
+\chi\{x=\pm y\}
+\chi\{xy=\pm1\}
=
-16-4-4
=
-24.
\]

The value $|\chi(X)|=24$ gives the expected critical-point count for a generic logarithmic one-form on this very affine threefold \cite{Huh2013}. The physical twist, however, belongs to a specified six-parameter family, so genericity in the unrestricted space of logarithmic one-forms cannot simply be assumed. Our proof of the five-master result below therefore uses the critical quotient directly and does not require an a priori identification of the full twisted-cohomology dimension with $24$.

Writing the logarithmic one-form as
\[
\omega
=
W_x\,d\log x
+
W_y\,d\log y
+
W_z\,d\log z,
\]
the critical ideal in $\mathcal O(X_{K_{\rm phys}})$ is
\[
\mathcal I_{\rm crit}
=
\langle W_x,W_y,W_z\rangle .
\]
Its zero locus coincides with the critical locus of $\log K_f$ on $X$.


For explicit polynomial computations, we clear the denominators in the critical equations and saturate by the excluded divisor. This saturation removes components supported on divisors excluded from the configuration space $X$. The resulting critical quotient is
\[ \mathcal Q_{\rm crit}
=
\mathcal O(X_{K_{\rm phys}})/\mathcal I_{\rm crit}. \]
The generic critical locus is zero-dimensional for the physical six-parameter family. Indeed, the equations $W_x=W_y=0$ independently determine $\alpha$ and $a$, while
\[
W_z=-xy(z-z^{-1})
\left(\frac{b}{\Delta}+\frac{c}{\Sigma}\right).
\]
Away from $z=\pm1$, the last equation imposes an additional independent linear condition on the exponents. On $z=\pm1$, it vanishes identically, but the dimension of the configuration-space stratum decreases by one. Consequently, the critical-point incidence variety has dimension six, equal to the dimension of the physical parameter space. Its generic fibers are therefore zero-dimensional whenever nonempty.

To obtain the required lower bound, it is sufficient to exhibit five representatives in the $I/J$ cohomological span that are linearly independent in twisted cohomology. A convenient choice is
\[
\mathcal B_5=\{I_1,I_2,I_3,J_{13},J_{14}\}.
\]
This choice is not meant to define a canonical basis; rather, it provides a set of representatives for which independence can be tested explicitly in the critical quotient and then lifted to the physical twisted cohomology.

Using the logarithmic representatives introduced in section~\ref{sec:family}, we express the selected top-form representatives relative to the common
volume form
\[
\Omega_0=d\log x\wedge d\log y\wedge d\log z
\]
as
\[
\varphi_i=g_i\,\Omega_0 .
\]
Here $\varphi_i=i\phi_i$, with the common factor $1/i$ in the physical representatives of section~\ref{sec:family} suppressed. The corresponding coefficient functions are
\begin{equation}
g_1=1,\qquad
g_2=y\partial_y\log\frac{\Delta}{\Sigma},\qquad
g_3=x\partial_x\log\frac{\Delta}{\Sigma},\qquad
g_{13}=\frac{N_{13}}{\Delta\Sigma},\qquad
g_{14}=\frac{N_{14}}{\Delta\Sigma},
\label{eq:gis}
\end{equation}
where \(N_{13}\) and \(N_{14}\) are the numerators of the corresponding rational representatives, given explicitly in appendix \ref{app:J-rational}.
The problem of establishing a five-dimensional lower bound is therefore reduced to testing the independence of these five coefficient functions in the critical quotient.

As an exact algebraic check, we specialize the six physical parameters to $(2,3,5,7,11,13)$ over the finite field $\mathbb F_{1009}$. The saturated critical quotient has dimension $24$, and the five representatives in eq.~\eqref{eq:gis} have rank five. The rank-five result is also obtained for the same integer parameter tuple over the finite field $\mathbb F_{1013}$.

To establish generic independence over $K_{\rm phys}$, we additionally verify nondegeneracy at the specialization over $\mathbb F_{1009}$. Specifically, the ideal obtained by adjoining the Jacobian determinant of the three polynomial critical equations to the saturated critical ideal is the unit ideal. Thus the specialized critical scheme is finite and reduced,
and hence etale over $\mathbb F_{1009}$.

Its geometric points lift, by Hensel's lemma, to characteristic zero over a finite unramified extension of $\mathbb Q_{1009}$. Since the five functions are linearly independent in the reduced special fiber, their evaluations at five suitable geometric critical points have a nonzero determinant. The corresponding determinant remains nonzero after lifting, establishing independence at the characteristic-zero specialization.

Finally, the critical equations define an algebraic family over the six-dimensional parameter space. Near the lifted simple critical points, these points and the nonzero evaluation determinant persist under deformation of the parameters. The five functions are therefore linearly independent in the generic critical quotient over $K_{\rm phys}$.

The function-field basis theorem of Matsubara--Heo and Telen \cite{MatsubaraHeoTelen2025} assumes generic parameters, whereas the present exponents are restricted to the physical kinematic slice. We therefore work directly over $K_{\rm phys}$ and introduce the deformed differential
\begin{equation}
\nabla_\delta=\delta\,d+\omega\wedge .
\label{eq:delta}
\end{equation}

At $\delta=0$, the deformed complex becomes the Koszul complex $(\Omega^\bullet,\omega\wedge)$, whose top cohomology is $\mathcal Q_{\rm crit}\Omega_0$. Since the generic critical ideal is zero-dimensional in the regular three-dimensional coordinate ring of $X$, its three generators form a regular sequence. Consequently,
\[
H^k(\Omega^\bullet,\omega\wedge)=0,
\qquad k<3.
\]

Consider the $\delta$-adically completed complex
\[
C^\bullet=
\bigl(\Omega^\bullet[[\delta]],\nabla_\delta\bigr).
\]
Multiplication by $\delta$ gives a short exact sequence of complexes
\[
0\longrightarrow C^\bullet
\xrightarrow{\ \delta\ }C^\bullet
\longrightarrow
(\Omega^\bullet,\omega\wedge)
\longrightarrow0.
\]
Since the Koszul cohomology vanishes below degree three, the associated long exact sequence implies
\[
H^3(C^\bullet)/\delta H^3(C^\bullet)
\simeq \mathcal Q_{\rm crit}\Omega_0,
\]
and multiplication by $\delta$ on $H^3(C^\bullet)$ is injective.

Suppose that the five selected classes satisfy a nontrivial linear relation after inverting $\delta$. Clearing denominators and dividing by the largest common power of $\delta$, we obtain a relation with coefficients in $K_{\rm phys}[[\delta]]$, at least one of which has a nonzero constant term. Injectivity of multiplication by $\delta$ allows this relation to be considered in $H^3(C^\bullet)$. Reduction modulo $\delta$ then gives a nontrivial relation among the five representatives in $\mathcal Q_{\rm crit}$, contradicting their established independence.

Thus the five classes remain independent over $K_{\rm phys}((\delta))$. For $\delta\neq0$,
\[
\nabla_\delta
=\delta\bigl(d+\delta^{-1}\omega\wedge\bigr).
\]
Because $\omega$ is linear in the six kinematic exponents, the substitution
\[
(\alpha,a,e,d,b,c)
\longmapsto
\delta^{-1}(\alpha,a,e,d,b,c)
\]
is an automorphism of the rational function field after adjoining $\delta$. A relation among the five classes in the physical twisted cohomology would therefore induce a relation in the deformed complex, contrary to the independence just proved. Hence the five classes are independent over $K_{\rm phys}$ in the physical twisted cohomology.

Hence the five classes $[I_1],[I_2],[I_3],[J_{13}],[J_{14}]$ are linearly independent;
\[
\operatorname{rank}_{K_{\rm phys}}
\bigl\{
[I_1],[I_2],[I_3],[J_{13}],[J_{14}]
\bigr\}
=5 .
\]

This establishes the required lower bound. Combining it with the upper bound from section~\ref{sec:counting}, we obtain
\begin{equation}
5\leq\dim_{K_{\rm phys}}\cV_{14}
\leq14-\operatorname{rank}R_{\rm ver}=5.
\end{equation}
Therefore,
\begin{equation}
\dim_{K_{\rm phys}}\cV_{14}=5.
\label{eq:exact-five-masters}
\end{equation}
Consequently, the five classes
$[I_1]$, $[I_2]$, $[I_3]$,
$[J_{13}]$, and $[J_{14}]$
form a basis of $\cV_{14}$ at generic complexified physical kinematics. This statement concerns the selected $I/J$ cohomological span, not necessarily the full top twisted cohomology.

Rank--nullity now gives a further consequence. Consider the $K_{\rm phys}$-linear map
\[
T:K_{\rm phys}^{14}
\longrightarrow
H^3(X_{K_{\rm phys}},\nabla_\omega),
\qquad
e_i\longmapsto[\varphi_i],
\]
where the fourteen classes are ordered as in eq.~\eqref{eq:v14}. Since $\operatorname{im}T=\cV_{14}$ has dimension five,
\[
\dim_{K_{\rm phys}}\ker T=14-5=9.
\]
The seven independent Ward relations and the two pointwise identities span a nine-dimensional subspace of this kernel. They therefore exhaust all linear relations over $K_{\rm phys}$ among the fourteen selected cohomology classes.

\section{Extension to the \(K\)-family and closure of the enlarged system}
\label{sec:Kfamily}

The five-dimensional result established in the previous sections concerns the scalar integral sector generated by the $I/J$ representatives. A natural question is whether this dimension remains unchanged after including the additional $K_i$ representatives appearing in the three-closed-string D-brane disk-amplitude decomposition of \cite{MousaviVelni2018}.
This enlargement is nontrivial because the new integrals introduce rational structures not present in the $I/J$ sector.

The purpose of this section is to test the closure of the twisted cohomological description under this enlargement. We show that the additional Ward identities and pointwise relations do not generate a new cohomology direction. The only exceptional case is the representative $K_3$, whose double-pole structure requires a separate pole-lowering argument. After this reduction, the enlarged I/J/K family is shown to lie in the same selected cohomological span identified for the I/J sector.

\subsection{World-sheet representation and exchange involution}
\label{subsec:worldsheetrep}

The $K$-family is defined on the same gauge-fixed world-sheet configuration space and with the same Koba--Nielsen twist used for the $I/J$ sector. The difference lies in the rational representatives multiplying the common twisted measure, some of which contain higher-order poles and require a separate reduction analysis. 

We retain the world-sheet variables $x$, $y$, $z=e^{i\theta}$ and the quantities $\Delta$ and $\Sigma$ introduced in section~\ref{sec:family}.
For the $K$-family it is also convenient to introduce
\[
\Gamma:=x^2+y^2+2xy\cos\theta .
\]

We use the same logarithmic volume form $\Omega_0$. Relative to this volume form, a rational prefactor $R_i$ in the $K$-family is represented by the coefficient function
\begin{equation}
g_{K_i}=x^2y^2R_i,
\end{equation}
up to the common orientation factor $1/i$. The rational coefficient functions \(g_{K_i}\) are collected in appendix \ref{app:K-rational}. For the argument below, it is useful to single out
\begin{equation}
g_{K_3}
=
\frac{2\Gamma(x^2-y^2)}{\Delta^2},
\end{equation}
whose double pole along $\Delta=0$ is treated separately in section~\ref{sec:polek3}.

The enlarged family inherits the exchange involution associated with interchanging the two closed-string insertions. In the world-sheet and kinematic variables defined previously, it acts as
\begin{equation}
\mathcal E:
(x,y,z;s,t,p,q)
\longmapsto
(y,x,z^{-1};t,s,q,p),
\qquad
u,v\ \text{fixed}.
\label{UVfixed}
\end{equation}
Its action on the $K$-family is
\begin{equation}
K_1\leftrightarrow K_2,\qquad
K_3\mapsto-K_3,\qquad
K_4\leftrightarrow-K_5,\qquad
K_6\leftrightarrow K_9,\qquad
K_7\leftrightarrow K_{10},\qquad
K_8\leftrightarrow K_{11}.
\label{eq:K-exchange}
\end{equation}
These properties generate the exchanged Ward primitives and control the treatment of $K_3$ below.

\subsection{Rank of the enlarged Ward system}

We now express the Ward identities of the enlarged $I/J/K$ family in terms of the twist exponents and physical kinematic variables defined in eqs.~(\ref{eq:KN-exponents}) and~(\ref{mandelstum}), respectively. The enlarged system is therefore considered over the same physical parameter space as the $I/J$ sector.

The seven Ward identities displayed in eq.~(23) of
Ref.~\cite{MousaviVelni2018}, together with their
$2\leftrightarrow3$ exchange images, generate fourteen Ward equations
for the enlarged $I/J/K$ system. Treating their coefficients over the
physical kinematic field, the corresponding Ward matrix has generic
rank
\begin{equation}
\operatorname{rank} R_{\mathrm{Ward}}^{(IJK)}=13 .
\label{eq:IJK-Ward-rank}
\end{equation}

We label the seven Ward identities in eq.~(23) of
Ref.~\cite{MousaviVelni2018}, in the order in which they appear
there, by $\mathcal{C}_1,\ldots,\mathcal{C}_7$ and denote their $2\leftrightarrow3$ exchange images by
$\widetilde{\mathcal{C}}_1,\ldots,\widetilde{\mathcal{C}}_7$.
Thus, $\mathcal{C}_1,\ldots,\mathcal{C}_4$ are the four $I/J$-sector identities, while $\mathcal{C}_5,\mathcal{C}_6,\mathcal{C}_7$ are the three identities containing the new $K$-integrals.

The single rank deficiency of the resulting fourteen-equation Ward system is accounted for by the unique generic syzygy
\begin{equation}
-2s\,\mathcal{C}_1
-v\,\mathcal{C}_2
-u\,\mathcal{C}_3
-2q\,\mathcal{C}_4
+2t\,\widetilde{\mathcal{C}}_1
+v\,\widetilde{\mathcal{C}}_2
+u\,\widetilde{\mathcal{C}}_3
+2p\,\widetilde{\mathcal{C}}_4
=0.
\label{eq:Kwardsyzygy}
\end{equation}

Since neither $\mathcal{C}_5,\mathcal{C}_6,\mathcal{C}_7$ nor their exchange images occur in
eq.~\eqref{eq:Kwardsyzygy}, the six genuinely new
$K$-containing Ward identities add six independent directions to the
rank-seven $I/J$ Ward module.

Appending the two pointwise $I/J$ identities derived in section~\ref{sec:worldsheet} to the rank-thirteen enlarged Ward system gives a relation matrix of generic rank
\begin{equation}
\operatorname{rank} R_{\mathrm{ver}}^{(IJK)}=15 .
\label{eq:IJK-verified-rank}
\end{equation}
An exact nonvanishing maximal minor is
\begin{equation}
2048\,p^2q\,s^2t^2
(s-t)(s+t)(u-v)(u+v)(sv-tu),
\label{eq:IJK-rank15-minor}
\end{equation}
so the rank statement holds generically.

\subsection{Twisted primitives for the six new Ward identities}

Owing to the exchange involution defined in eq.~\eqref{UVfixed}, it is sufficient to
construct three primitives explicitly; the remaining three follow
by exchange. For a two-form written in the cyclic logarithmic basis,
\begin{equation}
 \eta=A\,d\log y\wedge d\log z+
 B\,d\log z\wedge d\log x+
 C\,d\log x\wedge d\log y,
\end{equation}
one has
\begin{align}
 \frac{d\eta}{\Omega_0}
 &=x\partial_xA+y\partial_yB+z\partial_zC,\nonumber\\
 \frac{\nabla_\omega\eta}{\Omega_0}
 &=x\partial_xA+y\partial_yB+z\partial_zC
   +W_xA+W_yB+W_zC .
   \label{eq:gradianetta}
\end{align}
 This representation separates the ordinary exterior derivative from the twist contribution and therefore allows closure and twisted exactness to be checked independently.

The first $K$-dependent Ward relation, $\mathcal{C}_5$, is
\begin{equation}
2sJ_{12}+v(K_6+2K_8)-2pK_4+uK_7=0 .
\label{eq:identityjk1}
\end{equation}
An explicit two-form primitive for this relation is
\[
\eta^{(K)}_1
=
A_1\,d\log y\wedge d\log z
+
C_1\,d\log x\wedge d\log y .
\]
The coefficients $A_1$ and $C_1$ are obtained by imposing simultaneously
the ordinary closure condition $d\eta^{(K)}_1=0$ and the requirement that the twist contribution reproduce the rational representative of the
corresponding Ward combination. A convenient solution is
\[
\begin{aligned}
A_1 &= g_{J_{12}}
=\frac{x(1+y^2)\bigl[2x(1+y^2)-y(1+x^2)(z+z^{-1})\bigr]}{\Delta\Sigma},\\
C_1 &= -\frac{xy(x^2-1)(1+y^2)(z-z^{-1})}{\Delta\Sigma}.
\end{aligned}
\]
The ordinary exterior derivative vanishes,
\[
x\partial_x A_1+z\partial_z C_1=0,
\qquad\text{hence}\qquad
d\eta^{(K)}_1=0.
\]
Direct substitution into eq.~\eqref{eq:gradianetta} gives
\begin{equation}
\nabla_\omega\eta^{(K)}_1
=
\left[
2s\,g_{J_{12}}
+v(g_{K_6}+2g_{K_8})
-2p\,g_{K_4}
+u\,g_{K_7}
\right]\Omega_0 .
\end{equation}

The same construction gives primitives for $\mathcal{C}_6$ and $\mathcal{C}_7$. Their coefficients and twisted differentials are given in appendix~\ref{app:K-Ward-primitives}. For a two-form of the type used here, the exchange involution~\eqref{UVfixed} acts by
\begin{equation}
\mathcal E^*
\left(
A\,d\log y\wedge d\log z
+
C\,d\log x\wedge d\log y
\right)
=
\mathcal E(A)\,d\log z\wedge d\log x
-
\mathcal E(C)\,d\log x\wedge d\log y .
\label{eq:exchange-pullback-two-form}
\end{equation}

Applying this pullback to the three primitives gives twisted-exact representatives for the three exchanged Ward identities.

\subsection{Four pointwise identities and isolation of \(K_3\)}
\label{subsec:fourpoint}

 Using the rational representatives collected in appendix \ref{app:rational-representatives}, consider the set
\[
\{I_1,I_2,I_3,J_{13},J_{14},K_1,\ldots,K_{11}\}.
\]
A symbolic nullspace computation
gives rank twelve and nullity four. A convenient basis for this four-dimensional
pointwise nullspace is
\begin{equation}
\begin{aligned}
I_2-K_1+K_7 &=0,\\
I_3+K_2-K_6-2K_8 &=0,\\
I_3-K_2+K_{10} &=0,\\
I_2+K_1-K_9-2K_{11} &=0.
\end{aligned}
\label{eq:K-pointwise-identities}
\end{equation}
Combining these four pointwise identities with the six $K$-dependent Ward relations constructed above gives a square linear system for the ten $K$-integrals other than $K_3$. We collect them in the vector
\[
{\bf K}_{\neg3}=
(K_1,K_2,K_4,K_5,K_6,K_7,K_8,K_9,K_{10},K_{11})^{\mathsf T}.
\]
Retaining only the $K$-dependent coefficients, and ordering the four pointwise identities before the six $K$-dependent Ward relations, the corresponding coefficient matrix is
\[
M_{\neg3}=
\left(
\begin{array}{rrrrrrrrrr}
-1&0&0&0&0&1&0&0&0&0\\
0&1&0&0&-1&0&-2&0&0&0\\
0&-1&0&0&0&0&0&0&1&0\\
1&0&0&0&0&0&0&-1&0&-2\\
0&0&-2p&0&v&u&2v&0&0&0\\
0&0&0&-2p&u&v&-2u&0&0&0\\
u&v&0&0&0&0&0&0&0&2p\\
0&0&0&2q&0&0&0&v&u&2v\\
0&0&2q&0&0&0&0&u&v&-2u\\
v&u&0&0&0&0&2q&0&0&0
\end{array}
\right).
\]
The determinant of this matrix factorizes as
\begin{equation}
\det M_{\neg3}
=16\,p q\,(u-v)(u+v)(p+q+2u)^2 .
\label{eq:detMnonK3}
\end{equation}
Therefore, $M_{\neg 3}$ has rank ten away from the algebraic divisor defined by the vanishing of the factors in eq.~\eqref{eq:detMnonK3}. 
Moreover, the $K_3$ coefficient vanishes identically in all ten relations. Since \(M_{\neg 3}\) is generically invertible, the ten \(K_i\)-integrals with \(i\neq3\) reduce to the established \(I/J\) sector, leaving \(K_3\) as the only unresolved representative.

\subsection{Pole lowering and the final \(K_3\) relation}
\label{sec:polek3}
To lower the double pole \(\Delta^{-2}\) in \(g_{K_3}\), consider the two-form
\[
\eta_{\rm pl}
=
\frac{\Gamma}{(b-1)\Delta}
\left(
d\log y\wedge d\log z
-
d\log z\wedge d\log x
\right).
\]
The factor $b-1$ follows directly from the leading double-pole
coefficient: differentiating $\Delta^{-1}$ contributes $-1$, whereas
the $\Delta^b$ factor in the twist contributes $b$.

To evaluate the twisted differential of $\eta_{\rm pl}$, we use $W_x=x\partial_x\log K_f,\quad W_y=y\partial_y\log K_f,
$ which follow directly from the Koba–Nielsen factor in eq.\eqref{eq:Kf}:
\[
\begin{aligned}
W_x&=\alpha-\frac{2ex^2}{1-x^2}
+b\,\frac{x\partial_x\Delta}{\Delta}
+c\,\frac{x\partial_x\Sigma}{\Sigma},\\
W_y&=a-\frac{2dy^2}{1-y^2}
+b\,\frac{y\partial_y\Delta}{\Delta}
+c\,\frac{y\partial_y\Sigma}{\Sigma}.
\end{aligned}
\]
Substituting the expressions above into the twisted differential gives the exact
decomposition
\begin{equation}
g_{K_3}\Omega_0
=
\nabla_\omega\eta_{\rm pl}
+
R_{\rm pl}\Omega_0 ,
\label{eq:K3-pole-lowering}
\end{equation}
where
\begin{equation}
R_{\rm pl}
=
-\frac{1}{b-1}
\left[
\frac{2(x^2-y^2)}{\Delta}
+
\frac{\Gamma}{\Delta}
\left(
\alpha-a-\frac{2ex^2}{1-x^2}
+\frac{2dy^2}{1-y^2}
\right)
\right].
\label{eq:Rpl}
\end{equation}
The remainder \(R_{\rm pl}\) contains only a simple pole in \(\Delta\).
Using the physical kinematic relations and bringing the resulting
rational functions to a common denominator, one obtains the exact
pointwise identity
\begin{align}
(b-1)R_{\rm pl}
={}&
2(s-t)(g_J-g_{J_{13}})
+(p+q+1)(g_{I_2}-g_{I_3})
\nonumber\\
&+2p\,g_{I_7}-2q\,g_{I_4}
+(p-q)(g_{K_1}+g_{K_2})
+2p\,g_{K_4}+2q\,g_{K_5}.
\label{eq:Rpl-pointwise}
\end{align}

Combining eqs.~\eqref{eq:K3-pole-lowering} and
\eqref{eq:Rpl-pointwise}, and passing to twisted-cohomology classes,
gives the final relation
\begin{align}
0={}&(u+v-1)[K_3]
-2(s-t)([J]-[J_{13}])
-(p+q+1)([I_2]-[I_3])
\nonumber\\
&-2p[I_7]+2q[I_4]
-(p-q)([K_1]+[K_2])
-2p[K_4]-2q[K_5].
\label{eq:K3-final-relation}
\end{align}
As a consistency check, eq.~\eqref{eq:K3-final-relation} is odd under the exchange involution \eqref{UVfixed}: using the transformations in eq.~\eqref{eq:K-exchange} together with the corresponding exchange of kinematic invariants, each term changes sign.

\subsection{Five-master theorem}

\noindent\textbf{Theorem.}
For generic complexified physical kinematics,
\begin{equation}
\dim \operatorname{span}_{H^3}\{I/J/K\}=5.
\label{eq:IJK-five-dimensional}
\end{equation}
Using the five independent \(I/J\) classes established in section 6, a basis may be chosen as
\begin{equation}
\mathcal B_C=
\{[I_1],[I_2],[I_3],[J_{13}],[J_{14}]\}.
\label{eq:IJK-basis}
\end{equation}

\noindent\textit{Proof.}
Section \ref{sec:critical} establishes that the $I/J$ sector has dimension exactly five
and that the five classes displayed in eq.~\eqref{eq:IJK-basis} are
linearly independent. The rank-ten system constructed in section \ref{subsec:fourpoint} reduces every \(K_i\) with \(i\neq3\) to the \(I/J\) span.
Equation~\eqref{eq:K3-final-relation} reduces $K_3$ as well, provided that
\[
u+v-1\neq 0.
\]
Hence the enlarged \(I/J/K\) span is contained in the five-dimensional \(I/J\) span while containing the five independent classes in \(\mathcal B_C\), so its dimension is exactly five.

Here, “generic” refers to the intersection of the nonempty Zariski-open sets on which the relevant nonvanishing minor for the five \(I/J\) classes, the relevant Ward minors, the rank-ten reduction of the $K_{\neg 3}$ sector, and the coefficient $u+v-1$ are all nonzero. Where one of these nonvanishing conditions fails, the rank must be checked separately.

Before including eq.\eqref{eq:K3-final-relation}, the enlarged relation module has rank nineteen and vanishing \(K_3\) coefficient. Since eq.\eqref{eq:K3-final-relation} contains the generically nonzero coefficient \(u+v-1\) multiplying \([K_3]\), it is independent of the preceding relations, raising the generic rank to twenty. Thus the twenty-five representatives yield \(25-20=5\) master classes.

\subsection{Reduction of the scalar RR disk amplitude}
\label{subsec:physical-disk-reduction}

The five-master theorem has a direct consequence for the evaluation of scattering amplitudes. Consider the disk amplitude for an RR scalar, an antisymmetric state $B_2$, and a graviton $S_3$ on a D1-brane. The RNS expression of Becker et al.~\cite{BeckerBeckerRobbinsSu2016}, Eqs.~(3.22)--(3.24), contains 31 distinct integral labels when primed and unprimed labels are counted separately. We show how the same five scalar periods suffice for arbitrary polarizations, and then identify a physical component for which a single period suffices. The reduction starts with the four integrals whose integrands have a double pole: they occur in just two combinations. Denoting the periods of that reference by $\widetilde I_j$, radial integration by parts gives
\begin{equation}
\begin{aligned}
 U_- &\equiv (u+v)\widetilde I_{16}-2\widetilde I_{23}
      =-(s-t)\widetilde I_2+p\widetilde I_{20}-q\widetilde I'_{20},\\
 U_+ &\equiv 2\widetilde I_{24}-(u-v)\widetilde I_{17}
      =-(s+t)\widetilde I_1+p\widetilde I_{21}+q\widetilde I'_{21}.
\end{aligned}
\label{eq:physical-disk-pair}
\end{equation}
A prime exchanges the two disk variables in the rational integrand.
The two identities follow by integrating the difference and the sum,
respectively, of the logarithmic radial derivatives of the weighted
integrands for $\widetilde I_2$ and $\widetilde I_1$.
All remaining integrands are pointwise linear combinations of the
simple-pole representatives. The five-master theorem therefore gives the
full polarization-dependent amplitude, up to its common normalization, as
\begin{equation}
 \mathcal A_{C^{(0)}Bh}=T_1 I_1+T_2 I_2+T_3 I_3+T_4 J_{13}+T_5 J_{14}.
 \label{eq:physical-disk-five}
\end{equation}
For arbitrary physical polarizations and generic kinematics, we find that the five coefficients $T_\mu$ admit the common denominator $pq(p+q+2u)\,[2pq+3u(p+q)+4u^2]$.
Thus the factor $u+v-1$ encountered in reducing $K_3$ separately drops out of the coefficient denominators of the amplitude.

The shortening can be seen explicitly in a physical polarization component.
Let $P_2=Vp_2$, $p=P_2^2\ne0$, and choose a spacelike unit tangent vector
$\ell$ orthogonal to $P_2$. Take orthonormal vectors
$a,n$ normal to the brane with $n\cdot p_i=0$ for all three momenta,
$a\cdot p_1=a\cdot p_3=0$ and $a\cdot p_2\ne0$. Set
\[
 B_2=n\otimes\ell-\ell\otimes n,\qquad
 S_3=a\otimes n+n\otimes a,\qquad
 \mathcal P=(p_2Na)\epsilon_{ab}\ell^aP_2^b .
\]
\begin{samepage}
Both states are transverse and the graviton is traceless.
On the locus $s=t=0$, let $\rho=u/p$. The projection of the RNS expression
involves six scalar periods,
\begin{equation}
\begin{aligned}
 \frac{\mathcal A_{B_2S_3}}{\mathcal P}
  ={}&2(1+\rho)\big[U_-+U_+-(u+v)\widetilde I_{14}
                                 +(u-v)\widetilde I_{15}\big]\\
    &+4(q-u\rho)(\widetilde I_{14}-\widetilde I_{15}).
\end{aligned}
\label{eq:physical-disk-before}
\end{equation}
The pointwise identities $\widetilde I_{14}-\widetilde I_{15}=K_1$
and $\widetilde I_{14}+\widetilde I_{15}=K_2$, together with
Eq.~\eqref{eq:physical-disk-pair} and the Ward relations, reduce this to
\begin{equation}
 \mathcal A_{B_2S_3}=4(q+u)\,\mathcal P\,I_2.
 \label{eq:physical-disk-after}
\end{equation}
\end{samepage}
For the full amplitude, Eq.~\eqref{eq:physical-disk-five} separates the scalar integrations from the polarization dependence. Once the five master periods have been evaluated at a generic kinematic point, changing the external polarizations requires only a new contraction of the tensor coefficients $T_\mu$. For the polarizations chosen above and $s=t=0$, the six periods in Eq.~\eqref{eq:physical-disk-before} combine into the single $I_2$ term in Eq.~\eqref{eq:physical-disk-after}, with a nonzero coefficient for $q+u\ne0$. This further simplification holds even when the parallel momenta are noncollinear and have different squared norms.

\section{Discussion and conclusions}
\label{sec:discussion}

The central result of this work is a common world-sheet organization of relations that originally arose from spacetime gauge and T-duality constraints. In the $I/J$ sector, the Ward identities are represented by twisted-exact forms, while their unique generic syzygy admits a lift one degree lower in the same twisted de Rham complex.
Independent pointwise identities complete the relation structure of this sector. For the \(K\)-family, the six new Ward identities are generated by three explicit twisted primitives and their exchange images, while four additional relations hold pointwise at the level of the rational integrands.

For \(K_3\), the pole-lowering primitive constructed in section~\ref{sec:polek3} reduces the double pole to a simple-pole remainder, which is then reduced to the established \(I/J\) sector. Thus $K_3$ introduces no additional cohomology direction. Together with the independent five-class lower bound established by the critical-quotient analysis of section~\ref{sec:critical}, this proves that the enlarged $I/J/K$ family spans exactly five independent twisted-cohomology classes at generic complexified physical kinematics. This identifies a minimal cohomological set of scalar representatives for the $I/J/K$ decomposition of the disk amplitude.

In the D1 amplitude, combining the double-pole terms before reduction removes the factor $u+v-1$ from the master coefficients. For the transverse polarizations of section~\ref{subsec:physical-disk-reduction} at $s=t=0$, the six periods in eq.~\eqref{eq:physical-disk-before} reduce to the single $I_2$ contribution in eq.~\eqref{eq:physical-disk-after}.

The five-dimensional result holds at generic kinematics; resonant or rank-jumping loci may require separate analysis.

The reduction does not rely on numerical rank evidence alone. The explicit twisted primitives, the one-degree-lower lift of the Ward syzygy, the physical-slice degeneration, and the pole-lowering construction for $K_3$ provide explicit certificates for the principal steps of the reduction. Computer-algebra calculations support the exact rank and ideal computations, while the structural origin of the relations remains explicit at the level of the world-sheet integrands and their
twisted-cohomology classes.

The cohomological organization uncovered here suggests several
directions for future investigation. A natural question is whether analogous twisted complexes and reduction mechanisms persist for broader classes of D-brane disk integrals, including systems with additional insertions or less constrained kinematics. 

\section*{Acknowledgments}

The author acknowledges the use of ChatGPT (OpenAI) for assistance with language editing and formatting during the preparation of the manuscript.

\appendix

\section{Rational representatives of the \texorpdfstring{$I/J/K$}{I/J/K} family}
\label{app:rational-representatives}

This appendix records the rational representatives needed to reproduce the integrand-level identities, twisted-exact constructions, critical-quotient calculation, and the reduction of the enlarged $I/J/K$ family.  We use the
world-sheet variables and the functions $\Delta$, $\Sigma$, $Q=\Delta/\Sigma$, $\Gamma$, and $S=z+z^{-1}$ defined in the main text, together with the logarithmic volume form
\[
  \Omega_0=d\log x\wedge d\log y\wedge d\log z.
\]
For the \(I/J\) sector we use \(g_A=xyR_A\), while for the \(K\)-sector we retain the convention \(g_{K_i}=x^2y^2R_i\) of section \ref{subsec:worldsheetrep}. Thus the formulas below determine the corresponding original rational prefactors \(R_A\) and \(R_i\) uniquely.

For compactness, define
\[
\begin{aligned}
U_x &:=  2x(1+y^2)-y(1+x^2)S, &
U_y &:= 2y(1+x^2)-x(1+y^2)S,\\
\mathcal A &:= xyS, &
\mathcal D &:= x^2y^2(z-z^{-1})^2,\\
\mathcal P &:= x^2y^2S^2-x^4y^2-x^2y^4-x^2-y^2, &
N_0 &:= x^4y^2+x^2y^4-4x^2y^2+x^2+y^2.
\end{aligned}
\]

\subsection{\texorpdfstring{$I$}{I}-sector representatives}
\label{app:I-rational}

The five $I$-sector coefficients relative to $\Omega_0$ are
\[
\begin{aligned}
  g_{I_1} &= 1, \\
  g_{I_2} &=
  \frac{y(1-x^2)U_y}{\Delta\Sigma}, \\
  g_{I_3} &=
  \frac{x(1-y^2)U_x}{\Delta\Sigma}, \\
  g_{I_4} &= -\frac{1+y^2}{1-y^2}, \\
  g_{I_7} &= -\frac{1+x^2}{1-x^2}.
\end{aligned}
\]
The corresponding original prefactors follow from \(R_{I_a}=g_{I_a}/(xy)\). The exchange \(2\leftrightarrow3\) acts as \(R_{I_2}\leftrightarrow R_{I_3}\) and \(R_{I_4}\leftrightarrow R_{I_7}\).

\subsection{\texorpdfstring{$J$}{J}-sector representatives}
\label{app:J-rational}
The nine \(J\)-sector coefficients relative to \(\Omega_0\) are
\[
\begin{aligned}
  g_J &= \frac{\mathcal D}{\Delta\Sigma}, \\
  g_{J_1} &=
  \frac{x(1+x^2)(1-y^2)U_x}
       {(1-x^2)\Delta\Sigma}, \\
  g_{J_2} &=
  \frac{y(1+x^2)U_y}{\Delta\Sigma}, \\
  g_{J_3} &=
  -\frac{(1+x^2)(1+y^2)}{(1-x^2)(1-y^2)}, \\
  g_{J_4} &=
  \frac{y(1+y^2)(1-x^2)U_y}
       {(1-y^2)\Delta\Sigma}, \\
  g_{J_5} &=
  \frac{(1-x^2y^2)(x^2-y^2)}{\Delta\Sigma}, \\
  g_{J_{12}} &=
  \frac{x(1+y^2)U_x}{\Delta\Sigma}, \\
  g_{J_{13}} &= \frac{N_{13}}{\Delta\Sigma}, \\
  g_{J_{14}} &= \frac{N_{14}}{\Delta\Sigma},
\end{aligned}
\]
where the numerators entering the critical-quotient calculation of section \ref{sec:critical} are
\[
\begin{aligned}
 N_{13} &=
N_0+\mathcal A(1-x^2)(1-y^2), \\
N_{14} &=
N_0-\mathcal A(1-x^2)(1-y^2).
\end{aligned}
\]
The exchange involution gives
\[
g_{J_1}\leftrightarrow g_{J_4},\qquad
g_{J_2}\leftrightarrow g_{J_{12}},\qquad
g_{J_5}\mapsto -g_{J_5},
\]
while $g_J, g_{J_3}, g_{J_{13}}, g_{J_{14}}$
are invariant.

As a check, the coefficients above satisfy the pointwise identities \eqref{eq:pointwise-1} and \eqref{eq:pointwise-2} identically:
\[
g_{J_{12}}-g_{J_2}-2g_{J_5}=0,
\qquad
-g_{I_1}+g_J+g_{J_2}+g_{J_5}=0.
\]

\subsection{\texorpdfstring{$K$}{K}-sector representatives}
\label{app:K-rational}

For the enlarged \(K\)-family, the coefficient functions relative to \(\Omega_0\) are
\[
\begin{aligned}
  g_{K_1}
  &=\frac{(1-x^2)\bigl[(1+y^2)\mathcal A
       +2y^2(1+x^2)\bigr]}{\Delta\Sigma}, \\
  g_{K_2}
  &=\frac{(1-y^2)\bigl[(1+x^2)\mathcal A
       +2x^2(1+y^2)\bigr]}{\Delta\Sigma}, \\
  g_{K_3}
  &=\frac{2\Gamma(x^2-y^2)}{\Delta^2}, \\
  g_{K_4}
  &=\frac{(1+x^2)(1+y^2)
       \bigl[2x^2(1+y^2)-(1+x^2)\mathcal A\bigr]}
       {(1-x^2)\Delta\Sigma}, \\
  g_{K_5}
  &=\frac{(1+x^2)(1+y^2)
       \bigl[(1+y^2)\mathcal A-2y^2(1+x^2)\bigr]}
       {(1-y^2)\Delta\Sigma}, \\
  g_{K_6}
  &=-\frac{2(1+y^2)\mathcal P}
       {(1-y^2)\Delta\Sigma}, \\
  g_{K_7}
  &=\frac{2(1-x^2)(1+y^2)\mathcal A}
       {\Delta\Sigma}, \\
  g_{K_8}
  &=\frac{(1+y^2)
       \bigl[(x^2-y^2)(1-x^2y^2)+\mathcal D\bigr]}
       {(1-y^2)\Delta\Sigma}, \\
  g_{K_9}
  &=-\frac{2(1+x^2)\mathcal P}
       {(1-x^2)\Delta\Sigma}, \\
  g_{K_{10}}
  &=\frac{2(1+x^2)(1-y^2)\mathcal A}
       {\Delta\Sigma}, \\
  g_{K_{11}}
  &=\frac{(1+x^2)
       \bigl[(x^2-y^2)(-1+x^2y^2)+\mathcal D\bigr]}
       {(1-x^2)\Delta\Sigma}.
\end{aligned}
\]
The exchange involution of section~\ref{subsec:worldsheetrep} is manifest in these expressions:
\[
\begin{aligned}
g_{K_1}&\leftrightarrow g_{K_2},&
g_{K_3}&\mapsto -g_{K_3},&
g_{K_4}&\leftrightarrow -g_{K_5},\\
g_{K_6}&\leftrightarrow g_{K_9},&
g_{K_7}&\leftrightarrow g_{K_{10}},&
g_{K_8}&\leftrightarrow g_{K_{11}} .
\end{aligned}
\]
As a check on the normalization of the \(K\)-sector coefficients, the four pointwise identities used in section \ref{subsec:fourpoint} are satisfied identically:
\[
\begin{aligned}
  g_{I_2}-g_{K_1}+g_{K_7}&=0, \\
  g_{I_3}+g_{K_2}-g_{K_6}-2g_{K_8}&=0, \\
  g_{I_3}-g_{K_2}+g_{K_{10}}&=0, \\
  g_{I_2}+g_{K_1}-g_{K_9}-2g_{K_{11}}&=0.
\end{aligned}
\]
These identities provide algebraic checks of the normalization and exchange conventions.

\subsection{Primitives for the remaining
  \texorpdfstring{$K$}{K}-sector Ward identities}
\label{app:K-Ward-primitives}
The second $K$-dependent Ward relation, $\mathcal{C}_6$, is
\begin{equation}
-2sJ_4+vK_7-2pK_5+u(K_6-2K_8)=0 .
\end{equation}
Solving the same closure and matching conditions gives the two-form
\[
\eta^{(K)}_2
=
A_2\,d\log y\wedge d\log z
+
C_2\,d\log x\wedge d\log y ,
\]

where
\[
\begin{aligned}
A_2&=-g_{J_4},\\
C_2&=
-\frac{
xy(1+y^2)(z^2-1)
\bigl[
x^2y^2z+x^2z-2xyz^2-2xy+y^2z+z
\bigr]
}{
(x-yz)(y^2-1)(xy-z)(xz-y)(xyz-1)
}.
\end{aligned}
\]
As for the first primitive, one verifies
\[
x\partial_x A_2+z\partial_z C_2=0,
\qquad\qquad
d\eta^{(K)}_2=0.
\]
The corresponding twisted differential is therefore
\begin{equation}
\nabla_\omega\eta^{(K)}_2
=
\left[
-2s\,g_{J_4}
+v\,g_{K_7}
-2p\,g_{K_5}
+u\bigl(g_{K_6}-2g_{K_8}\bigr)
\right]\Omega_0 .
\end{equation}

The third $K$-dependent Ward relation, $\mathcal{C}_7$, is
\begin{equation}
2s(-J+J_5)+vK_2+2pK_{11}+uK_1=0 .
\label{eq:K-Ward-3}
\end{equation}
To express $J$ and $J_5$ in the common $\Omega_0$ representation,
we use the pointwise identities in eqs.~\eqref{eq:pointwise-1} and \eqref{eq:pointwise-2},
which give
\[
g_{J_5}=\frac{g_{J_{12}}-g_{J_2}}{2},
\qquad
g_J=1-g_{J_2}-g_{J_5}.
\label{eq:J-J5-coefficients}
\]
Solving the same closure and matching conditions as above, an explicit
two-form primitive may be chosen as
\[
\eta^{(K)}_3
=
A_3\,d\log y\wedge d\log z
+
C_3\,d\log x\wedge d\log y ,
\label{eq:K-primitive-3}
\]
with
\[
\begin{aligned}
A_3&=-g_J+g_{J_5},\\
C_3&=
-\frac{
xyz(x^2-1)(1+y^2)(z^2-1)
}{
(x-yz)(xy-z)(xz-y)(xyz-1)
}.
\end{aligned}
\]
As in the preceding two cases, the ordinary exterior derivative vanishes,
\[
d\eta^{(K)}_3=0.
\]
Applying the twisted differential then gives
\begin{equation}
\nabla_\omega\eta^{(K)}_3
=
\left[
2s(-g_J+g_{J_5})
+v\,g_{K_2}
+2p\,g_{K_{11}}
+u\,g_{K_1}
\right]\Omega_0 .
\label{eq:K-primitive-3-exact}
\end{equation}

Applying the pullback~\eqref{eq:exchange-pullback-two-form} to $\eta^{(K)}_1$, $\eta^{(K)}_2$ and $\eta^{(K)}_3$ gives
\begin{equation}
\begin{aligned}
\nabla_\omega\!\left(\mathcal E^*\eta^{(K)}_1\right)
&=
\left[
2t\,g_{J_2}
+v\left(g_{K_9}+2g_{K_{11}}\right)
+2q\,g_{K_5}
+u\,g_{K_{10}}
\right]\Omega_0,\\
\nabla_\omega\!\left(\mathcal E^*\eta^{(K)}_2\right)
&=
\left[
-2t\,g_{J_1}
+v\,g_{K_{10}}
+2q\,g_{K_4}
+u\left(g_{K_9}-2g_{K_{11}}\right)
\right]\Omega_0,\\
\nabla_\omega\!\left(\mathcal E^*\eta^{(K)}_3\right)
&=
\left[
2t\left(-g_J-g_{J_5}\right)
+v\,g_{K_1}
+2q\,g_{K_8}
+u\,g_{K_2}
\right]\Omega_0.
\end{aligned}
\end{equation}
Thus all six new $K$-dependent Ward identities admit explicit twisted-exact representatives.

\end{document}